\documentstyle{article}

\catcode`\@=11

\@addtoreset{equation}{section}
\catcode`\@=12
\def\sgn{\mathop{\rm sgn}\nolimits}
\begin{document}

\title{On time evolutions associated with the nonstationary Schr\"{o}dinger
equation}
\author{A.~K.~Pogrebkov \\
Steklov Mathematical Institute, \\
Gubkin St. 8, Moscow, 117966, GSP-1, Russia \\
e-mail: pogreb@mi.ras.ru}
\date{\today}
\maketitle

\begin{abstract}
The set of integrable symmetries of the nonstationary Schr\"{o}dinger
equation is shown to admit a natural decomposition into subsets of mutually
commuting symmetries. Hierarchies of time evolutions associated with each of
these subsets ultimately lead to nonlinear (possibly, operator) equations of
the Kadomtsev--Petviashvili I type or its higher analogues, thus
demonstrating that the linear problem itself constructively determines the
associated nonlinear integrable evolution equations and their hierarchies.
\end{abstract}

\tableofcontents

\pagebreak

\section{Introduction}

In this article, we use the nonstationary Schr\"{o}dinger~(NS) equation as an
example to demonstrate that a given linear problem determines the nonlinear
evolutions associated with it. Our approach is based on the study of the set
of symmetries for the NS equation using the inverse scattering transform
(IST). We here consider those symmetries that can be called integrable, which
means that the associated flows can be expressed in terms of the potential of
the linear problem or in terms of some special values of the Jost solutions.
The set of all integrable symmetries is decomposed into subsets of mutually
commuting ones; with each of these subsets, we associate a hierarchy of
evolution equations. We show that all these subsets of commuting symmetries
have a natural hierarchy structure, and we investigate the corresponding
evolutions. We find that the Jost solution of the original NS equation with
respect to the two lowest times of any such hierarchy also obeys an NS
equation with an operator-valued potential. Moreover, the third (or any
higher) time of the corresponding hierarchy provides an operator differential
equation whose linear part coincides with the linear part of the
Kadomtsev--Petviashvili~I (KPI) equation~\cite{KP} (or of its higher analogue
respectively). This linearized KPI equation naturally appears as the equation
for the spectral data here; it follows from a simple commutator identity that
the spectral data satisfy this equation (or its higher analogues) regardless
of the choice of the subclass of commuting symmetries. We also show that
there are two distinguished subclasses of commuting flows characterized by
the property that the above-mentioned operator potential is just a
multiplication operator. These subsets of symmetries correspond to the
standard evolutions and the ones that can be considered dual to them. We
consider these subsets in detail and show that the second one leads to a
transformation of both the dependent and independent variables, which
essentially extends the range of applicability of the IST.

The study of different symmetries associated with the NS equation (or other
linear problems) goes back to the dressing procedure~\cite{ZSh1,ZSh2} and was
investigated either by these methods and their extensions or in formal
algebraic terms (see~\cite{Z90}--\cite{K}). The approach adopted below is
peculiar: we work strictly in the IST framework~\cite{ZManakov}--\cite{BLP},
starting with the NS equation with respect to ($x_{1},x_{2}$) variables with
a real potential $u(x_{1},x_{2})$ assumed to be a smooth function that
rapidly decays at infinity and satisfies some weak conditions on its norm
\cite{Xin} to guaranty the solvability of the direct and inverse problems.

Our approach is based on a special version of the IST, the so-called
resolvent approach (see~\cite{first}--\cite{KPtmf}). In the next section, we
briefly describe some constructions that enable us to combine the 
methods of the IST with the algebraic schemes of the type in
\cite{DJKM}--\cite{Orlov}. To control the transformations of all the involved
objects---spectral data, Jost solutions, and the potential---we use the
so-called scattering theory on a nontrivial background~\cite{physicad}. In
Sec.~3, we consider integrable symmetries in this framework and derive
general results related to their commuting subsets (evolutions). The simplest
examples of such evolutions are given in Sec. 4.

\section{Extended operators and resolvent approach}

The first step in the resolvent approach consists in a special extension of
the differential operators. Precisely, any given differential operator $A$
with the kernel
\begin{equation}
A(x,x')=A(x_{1},x_{2},\partial _{x_{1}},\partial _{x_{2}})\delta
(x_{1}-x_{1}')\delta (x_{2}-x_{2}')  \label{kern}
\end{equation}
is replaced with the differential operator $A(q)$ with the kernel
\begin{equation}
A(x,x';q)=e_{}^{-q(x-x')}A(x,x')\equiv
A(x_{1},x_{2},\partial _{x_{1}}+q_{1},\partial _{x_{2}}+q_{2})\delta
(x-x'),  \label{1'}
\end{equation}
where $qx=q_{1}x_{1}+q_{2}x_{2}$ (we consider only the two-dimensional case
here). In~(\ref{1'}), $q$ is a real (two-component) vector; below, we often
omit explicit indication of the operator dependence on the variable $q$. The
standard definitions of the dual operator, the complex conjugate operator,
and the Hermitian conjugate operator are modified as follows:
\begin{eqnarray}
&&A^{{\rm dual}}(x,x';q)=A^{{\rm t}}(x',x;-q),\qquad
A^{*}(x,x';q)=\overline{A(x,x';q)},  \nonumber \\
&&A^{\dagger }(x,x';q)=\overline{A^{{\rm t}}(x',x;-q)},
\label{3a}
\end{eqnarray}
where the superscript t denotes transposition in the matrix case. In what
follows, we refer to the kernel $A(x,x';q)$ of the operator $A$ as the
kernel in the $x$ representation to distinguish it from the kernel of the
same operator in the ($p,{\bf q}$) representation defined via the ``shifted''
Fourier transform,
\begin{equation}
A(p;{\bf q})=\frac{1}{(2\pi )^{2}}\int dx\int dx'e_{}^{i(p+{\bf q}
_{\Re })x-i{\bf q}_{\Re }x'}A(x,x';{\bf q}_{\Im }),
\label{2a}
\end{equation}
where $p$ and ${\bf q}$ are respectively real and complex two-dimensional
vectors. In these terms, the meaning of this extension of the kernels of
differential operators can be clarified by observing that the kernels
$A(p;{\bf q})$ of differential operators are polynomial functions of the
variable ${\bf q}$. As shown in~\cite{first}, this analyticity property
of the extended differential operators considerably simplifies the study of
their spectral transform.

Below, we consider integral operators $A(q)$ ($A,$ for short) with kernels
$A(x,x';q)$ belonging to the space ${\cal S}'$ of the Schwartz distributions
with respect to all six real variables, $x_{1}$, $x_{2}$, $x_{1}'$, $x_{2}'$,
$q_{1}$, and $q_{2}$ (or in the space of the ($p,{\bf q}$) variables with
respect to~(\ref{2a})). We also consider a special subclass ${\cal M}\subset
{\cal S}'$ of such operators, whose kernels are Schwartz distributions with
respect to the variables $x$ and $x'$ and are piecewise continuous functions
with respect to the variable $q$.

The relevance of this subclass is demonstrated by the following simple
example. Let $D_{j}$ denote the extension of the differential operator
$\partial _{x_{j}}$, $j=1,2$, by~(\ref{1'}), i.e.,
\begin{equation}
D_{j}(x,x';q)=(\partial _{x_{j}}+q_{j})\delta (x-x'),\quad j=1,2.
\label{10}
\end{equation}
This operator is uniquely invertible on the subclass ${\cal M}$. Its inverse
is given by the kernel
\begin{equation}
D_{j}^{-1}(x,x';q)=\sgn q_{j}e^{-q_{j}(x_{j}-x_{j}')}\theta
(q_{j}^{}(x_{j}^{}-x_{j}'))\delta (x_{j+1}^{}-x'_{j+1}),\quad j=1,2,
\label{10'}
\end{equation}
($j+1$ is understood mod 2) and is exactly the standard resolvent of the
operator $\partial _{x_{j}}$. This observation gave the name {\sl resolvent
approach} for the method based on the above extension of the original
differential operators. The uniqueness of inverse~(\ref{10'}) follows because
the only (up to a factor) annulator of $D_{j}$ belonging to ${\cal S}'$ has
the kernel $\delta (q_{j}^{})\delta(x_{j+1}^{}-x_{j+1}')$, which does not
belong to ${\cal M}$. We also mention that if the $q$ dependence of the
extended differential operator can be trivially removed by multiplying the
kernel by $\exp q(x-x')$ (cf.~(\ref{1'})), the $q$ dependence of the inverse
operator cannot be removed in this way, as is shown by the example in
(\ref{10'}).

We turn to the NS operator
\begin{equation}
{\cal L}_{x}^{}=i\partial _{x_{2}}^{}+\partial
_{x_{1}}^{2}-u(x_{1},x_{2}).  \label{7}
\end{equation}
Using~(\ref{1'}), we can write its extension as
\begin{equation}
L=L_{0}^{}-U,  \label{12}
\end{equation}
where $L_{0}$ is the extension corresponding to the zero-potential case
and is equal to
\begin{equation}
L_{0}^{}=iD_{2}^{}+D_{1}^{2}  \label{13}
\end{equation}
by~(\ref{10}). In~(\ref{12}), we introduce the multiplication operator $U$
with the kernel
\begin{equation}
U(x,x';q)=u(x)\delta (x-x').  \label{a11}
\end{equation}

The main object in our approach is the (extended) resolvent of  operator
(\ref{12}) defined as the inverse of the operator $L$ on the subclass
${\cal M}$, i.e.,
\begin{equation}
LM=ML=I,  \label{14}
\end{equation}
where $I$ is the unity operator,
\begin{equation}
I(x,x';q)=\delta (x-x').  \label{15}
\end{equation}
We mention that the kernel of the resolvent $M_{0}$ of the zero-potential
operator $L_{0}$~(\ref{13}) is equal to
\begin{equation}
M_{0}(x,x';q)=\frac{\sgn (x_{2}-x_{2}')}{2\pi i}\int d\alpha
\theta \Bigl((q_{2}-2\alpha q_{1})(x_{2}-x'_{2})\Bigr)
e_{}^{-[q+i\ell (\alpha +iq_{1})](x-x')},  \label{16}
\end{equation}
where we introduce the special two-vector
\begin{equation}
\ell (\alpha )=(\alpha ,\alpha _{}^{2}).  \label{l}
\end{equation}
Therefore, this resolvent indeed belongs to ${\cal M},$ and it is easy to see
that it is unique in this subclass. In general, the existence of an inverse
in the space of Schwartz distributions with respect to the variables $x$ and
$x'$ follows from the H\"{o}rmander theory, but the necessary properties with
respect to the variable $q$, as well as the uniqueness statement, must be
proved. This goes beyond the frame of this paper. We therefore assume here
that the potential $u(x)$ is a smooth function rapidly decaying at infinity;
the resolvent therefore exists and is unique in ${\cal M}$. The class of such
potentials is not empty because imposing the same small norm assumption used
in~\cite{Xin} is sufficient to prove the existence of the Jost solutions.
Moreover, we know (see~\cite{total}) that the resolvent in fact exists in
much more complicated situations, for instance, when the potential has a
nontrivial behavior at infinity.

To demonstrate that the extended resolvent generalizes all notions of the
standard approach, we mention that it is directly related to the Green's
function of operator~(\ref{7}). Indeed, if we introduce
\begin{equation}
G(x,x',{\bf k})=e_{}^{(x-x')\ell _{\Im }({\bf k})}M(x,x';
\ell _{\Im }^{}({\bf k})),  \label{175}
\end{equation}
then it is easy to verify that
\begin{equation}
{\cal L}_{x}^{}G(x,x';{\bf k})=\delta (x-x').
\label{176}
\end{equation}

We already mentioned that $L(p;{\bf q})$ is an analytic (polynomial) function
of ${\bf q}$. Taking into account that the standard product of any two
operators in terms of the ($p,{\bf q}$) variables takes the form
(cf.~(\ref{2a}))
\begin{equation}
(AB)(p;{\bf q})=\int dp'A(p-p';{\bf q}+p')B(p';{\bf q}),  \label{180}
\end{equation}
we see that $M(p;{\bf q})$ must in some sense be close to an analytic
function of the variable ${\bf q}$. As shown in~\cite{first,total}, studying
its lack of analyticity leads to the notion of the Jost solution. To be more
precise, we introduce the operator $\bar{\partial}_{2}M_{0}$ whose kernel in
the ($p,{\bf q}$) representation is equal to $\partial M_{0}(p;{\bf q})/
\partial {\bf \bar{q}}_{2}$. Then it is easy to see that
\begin{equation}
(\bar{\partial}_{2}M_{0})(x,x';q)=\frac{1}{8\pi |q_{1}|}\exp
\!\left\{ -i\ell _{\Re }^{}\Bigl(\frac{q_{2}}{2q_{1}}+iq_{1}^{}\Bigr)
(x-x')\right\}.  \label{17}
\end{equation}
It was shown in~\cite{total} that the lack of analyticity of the resolvent
$M$ is given in terms of two special reductions of the (truncated) resolvent
itself. These operators are defined as
\begin{eqnarray}
&&|\nu \rangle (x,x';q)=2\delta (x_{2}^{}-x_{2}')\int
dq_{2}^{}\Bigl((ML_{0}^{})\bar{\partial}_{2}^{}M_{0}^{}\Bigr)(x,x';q),
\label{181} \\
&&\langle \omega |(x,x';q)=2\delta (x_{2}-x_{2}')\int
dq_{2}^{}\Bigl((\bar{\partial}_{2}^{}M_{0}^{})(L_{0}^{}M)\Bigr)
(x,x';q).  \label{182}
\end{eqnarray}
That they are reduced values is more explicit in the ($p,{\bf q}$)
representation: for example,
$|\nu \rangle (p,{\bf q})=(ML_{0})(p,{\bf q})|_{{\bf q}_{2}={\bf q}_{1}^{2}}$.
Both kernels $|\nu \rangle (p,{\bf q})$ and $\langle \omega |(p,{\bf q})$ are
independent of ${\bf q}_{2}$ and are analytic functions of ${\bf q}_{1}$ in
the upper and lower half-planes with a possible discontinuity at the real
${\bf q}_{1}$ axis. Both tend to $\delta(p)$ as 
${\bf q}_{1}\rightarrow\infty$. We call these operators the Jost solutions 
of the NS equation (and its dual), because if we introduce the functions
\begin{eqnarray}
&&\chi (x,{\bf k})=\int dx'e_{}^{i{\bf k}_{\Re}(x_{1}-
x_{1}')}|\nu \rangle (x,x';{\bf k}_{\Im }),
\label{23c} \\
&&\xi (x',{\bf k})=\int dxe_{}^{i{\bf k}_{\Re}(x_{1}-
x_{1}')}\langle \omega |(x,x';{\bf k}_{\Im })
\label{23d}
\end{eqnarray}
for some complex ${\bf k}$, then the standard Jost solutions are given by
\begin{equation}
\Phi (x,{\bf k})=e_{}^{-i\ell ({\bf k})x}\chi (x;{\bf k}),\qquad
\Psi (x,{\bf k})=e_{}^{i\ell ({\bf k})x}\xi (x;{\bf k}).  \label{20}
\end{equation}
We emphasize that $\Phi$ and $\Psi$ are the Jost solutions of the original
operator~(\ref{7}), not of the extended one, i.e., they satisfy
\begin{equation}
{\cal L}_{x}^{}\Phi (x,{\bf k})=0,\qquad {\cal L}_{x}^{{\rm dual}}
\Psi (x,{\bf k})=0,  \label{201}
\end{equation}
while the extension of the operator $L$ causes the Jost solutions to be
determined by reductions~(\ref{181}) and~(\ref{182}) as functions of the
complex spectral parameter ${\bf k}$. Precisely,
\begin{equation}
i\partial _{x_{2}}^{}\chi (x,{\bf k})+\partial _{x_{1}}^{2}\chi (x,{\bf k})
=u(x)\chi (x,{\bf k})+2i{\bf k}\partial _{x_{1}}^{}\chi (x,{\bf k}),
\label{2011}
\end{equation}
and it is easy to verify that by the above definitions, $\chi (x,{\bf k})$ is
normalized to 1 at ${\bf k}\rightarrow\infty$.

Operators~(\ref{181}) and~(\ref{182}) satisfy the ``differential equations"
\begin{equation}
L|\nu \rangle =|\nu \rangle L_{0},\qquad \langle \omega |L=L_{0}\langle
\omega |,  \label{23}
\end{equation}
which are of course equivalent to Eqs. (\ref{201}) after transformations
(\ref{23c}), (\ref{23d}), and (\ref{20}). The reality condition for the
potential $u(x)$ is equivalent to
\begin{equation}
|\nu \rangle _{}^{\dagger }=\langle \omega |,  \label{2302}
\end{equation}
which can also be written as $\overline{\Phi(x,{\bf k})}=\Psi(x,\bar{\bf k})$.

In~\cite{KPJMP}, we also proved that the operators $|\nu\rangle$ and $\langle
\omega |$ are the inverses of each other,
\begin{equation}
\langle \omega |\nu \rangle =I,\qquad |\nu \rangle \langle \omega |=I,
\label{2326}
\end{equation}
and that we have a bilinear representation of the operator $L$ and of the
resolvent $M$ by their means,
\begin{eqnarray}
&&L=|\nu \rangle L_{0}\langle \omega |,  \label{24} \\
&&M=|\nu \rangle M_{0}\langle \omega |.  \label{25}
\end{eqnarray}

Using the resolvent approach, we thus obtain equations that have the explicit
meaning of dressing the zero-potential operator $L_{0}$ with the operators
$|\nu\rangle$ and $\langle\omega|$. One feature of our approach is that these
dressing operators are in turn given as reductions~(\ref{181}) and
(\ref{182}) of the resolvent itself.

To formulate the inverse problem, we use the analyticity properties of the
kernels $|\nu\rangle(p,{\bf q})$ and $\langle\omega|(p,{\bf q})$ and
introduce their boundary values on the real axis from the upper and lower
half-planes. In terms of the ($x,q$) representation, the corresponding kernels
are defined by means of the limits
\begin{eqnarray}
&&|\nu ^{\pm }\rangle (x,x';q)=\lim_{q_{1}\rightarrow \pm 0}|\nu
\rangle (x,x';q),  \label{nu:pm} \\
&&\langle \omega ^{\pm }|(x,x';q)=\lim_{q_{1}\rightarrow \pm
0}\langle \omega |(x,x';q),  \label{o:pm}
\end{eqnarray}
where the l.h.s.'s, of course, do not depend on $q$. Because the potential is
real, they are mutually conjugate,
\begin{equation}
|\nu ^{\pm }\rangle ^{\dagger }=\langle \omega ^{\mp }|  \label{herm:Jpm}
\end{equation}
(see~(\ref{3a}) and~(\ref{2302})). The spectral data are defined~\cite
{kpshort,KPJMP} as
\begin{equation}
F=\langle \omega ^{-}|\nu ^{+}\rangle ,  \label{sd}
\end{equation}
and by~(\ref{2326}),
\begin{equation}
|\nu ^{+}\rangle =|\nu ^{-}\rangle F,\qquad \langle \omega ^{-}|=F\langle
\omega ^{+}|.  \label{spJ}
\end{equation}
These spectral data have the properties
\begin{equation}
F^{\dagger }=F,\qquad [L_{0},F]=0,\qquad q=0.  \label{F:2}
\end{equation}
In terms of the ($p,{\bf q}$) representation, the last equality means that
there exists a function $f$ of two real variables such that we have the
representation
\begin{equation}
F(p;{\bf q})=\delta (p_{2}-p_{1}(p_{1}+2{\bf q}_{1\Re }))f(p_{1}+
{\bf q}_{1\Re };{\bf q}_{1\Re }),  \label{sp}
\end{equation}
and then~(\ref{F:2}) means that $\overline{f(\alpha,\beta)}=f(\beta,\alpha)$.
In these terms, we can use~(\ref{23c}) to rewrite the first equation in
(\ref{spJ}), for instance, in the form
\begin{equation}
\chi _{}^{+}(x,k)=\int d\alpha \chi _{}^{-}(x,\alpha )e_{}^{ix(\ell
(k)-\ell (\alpha ))}f(\alpha ,k).  \label{spJ1}
\end{equation}
We thus see that Eqs.~(\ref{spJ}) are equivalent to the standard (see
\cite{ZManakov,Manakov}) formulation of the nonlocal Riemann--Hilbert problem
for the Jost solutions of the NS equation. In what follows, we assume the
unique solvability of these equations.

\section{Integrable symmetries and time evolutions}

\subsection{Deformations of the spectral data}

We now consider two potentials: $u(x)$ with the corresponding operators $L$,
$M$, $|\nu\rangle$, and $\langle\omega|$ and the spectral data $F$ and the
potential $\tilde{u}(x)$ with the corresponding $\widetilde{L}$,
$\widetilde{M}$, $|\tilde{\nu}\rangle$, $\langle\tilde{\omega}|$, and
$\widetilde{F}$. Then the two resolvents are related via~(\ref{2326}) and
(\ref{25}) using ``scattering on a nontrivial background"~\cite{physicad},
\begin{equation}
\widetilde{M}=\eta M\eta ^{-1},  \label{tMM}
\end{equation}
where we introduce an operator $\eta$ such that
\begin{equation}
\eta =|\widetilde{\nu }\rangle \langle \omega |,\qquad \eta ^{-1}=|\nu
\rangle \langle \widetilde{\omega }|.  \label{etaJ}
\end{equation}
This operator is unitary; because of~(\ref{2302}),
\begin{equation}
\eta ^{\dagger }=\eta ^{-1};  \label{unit}
\end{equation}
its kernel in the ($p,{\bf q}$) representation does not depend on
${\bf q}_{2}$, $\eta(p;{\bf q})=\eta (p;{\bf q}_{1})$; it is analytic for
${\bf q}_{1\Im}\neq0$ and satisfies the asymptotic condition
\begin{equation}
\lim_{{\bf q}_{1}{\bf \rightarrow \infty }}\eta (p;{\bf q}_{1})=\delta (p).
\label{as:eta}
\end{equation}
For the boundary values of $\eta$ on the real ${\bf q}_{1}$ axis (defined by
analogy with~(\ref{nu:pm}) and (\ref{o:pm})), we obtain
\begin{equation}
\eta ^{\pm }=|\widetilde{\nu }^{\,\pm }\rangle \langle \omega ^{\pm }|
\label{eta:pm}
\end{equation}
from~(\ref{etaJ}), and from~(\ref{spJ}), we then find
\begin{equation}
\eta ^{+}-\eta ^{-}=|\tilde{\nu}^{-}\rangle (\widetilde{F}-F)\langle \omega
^{+}|.  \label{disc:eta}
\end{equation}

We suppose that a curve in the spectral-data space is parametrized by $t$.
Correspondingly, all objects $M$, $L$, $U$, $|\nu\rangle$, and
$\langle\omega|$ become dependent on $t$; we therefore have a symmetry of the
NS equation. Let the nontilde variables correspond to some value $t$ and the
tilde variables correspond to the value $t+\tau $. Then $\eta=\eta(t,\tau)$,
and by~(\ref{2326}) and~(\ref{etaJ}), $\eta(t,0)=I$. We set
\begin{equation}
A=-i\frac{\partial \eta }{\partial \tau }\biggl|_{\tau =0}.  \label{def:A}
\end{equation}
This operator is Hermitian by~(\ref{unit}), $A^{\dagger }=A$, and its kernel
$A(p,{\bf q})$ is analytic in the upper and lower half-planes and by
(\ref{as:eta}) satisfies the asymptotic condition
\begin{equation}
\lim_{{\bf q}_{1}{\bf \rightarrow \infty }}A(p;{\bf q}_{1})=0.  \label{as:A}
\end{equation}
It follows from~(\ref{disc:eta}) that the boundary values of this operator at
the real axis satisfy
\begin{equation}
A_{}^{+}-A_{}^{-}=-i|\nu _{}^{-}\rangle \frac{\partial F}{\partial t}
\langle \omega _{}^{+}|.  \label{disc:A}
\end{equation}

It is clear that the analyticity properties of $A(p;{\bf q})$, together with
(\ref{as:A}) and~(\ref{disc:A}), uniquely determine the operator $A$ in the
($p,{\bf q}$) representation via the Cauchy formula
\begin{equation}
A(p;{\bf q})=\frac{-1}{2\pi }\int \frac{d{\bf q}_{1\Re }'}{{\bf q}
_{1\Re }'-{\bf q}_{1}}\left( |\nu _{}^{-}\rangle \frac{\partial F}
{\partial t}\langle \omega _{}^{+}|\right) \!(p;{\bf q}');
\label{C1}
\end{equation}
we recall that the kernels of all objects in the r.h.s., $|\nu\rangle$, $F$,
and $\langle\omega|$, depend neither on ${\bf q}_{2}$ nor on ${\bf q}_{1\Im}$
in the ($p,{\bf q}$) representation. We can use~(\ref{2a}) to rewrite this
kernel in the $x$ representation as
\begin{equation}
A(x,x';q)=-i\sgn q_{1}^{}
e_{}^{-q_{1}(x_{1}-x_{1}')}\theta
((x_{1}^{}-x_{1}')q_{1}^{})\left( |\nu _{}^{-}\rangle \frac{\partial F}
{\partial t}\langle \omega _{}^{+}|\right) \!(x,x';q),
\label{C2}
\end{equation}
where the last factor is independent of $q$ by construction.

If $A$ is known, then for the resolvent and the operator $L$ itself, we have
\begin{equation}
\frac{\partial M}{\partial t}=i[A,M],\qquad \frac{\partial L}{\partial t}
=i[A,L]  \label{der:M}
\end{equation}
from~(\ref{tMM}) and~(\ref{def:A}). Moreover, using~(\ref{2326}) to rewrite
(\ref{etaJ}) as $|\widetilde{\nu}\rangle=\eta|\nu\rangle$ and $\langle
\widetilde{\omega}|=\langle\omega|\eta^{-1}$, we obtain the derivatives of
the Jost solutions along the curve from~(\ref{def:A}):
\begin{equation}
\frac{\partial |\nu \rangle }{\partial t}=iA|\nu \rangle ,\qquad
\frac{\partial \langle \omega |}{\partial t}=-i\langle \omega |A.
\label{der:J}
\end{equation}

The flows of the operator $L$, the resolvent, and the Jost solutions are
determined using the IST and are given by the operator $A$, i.e., by dressing
the flow of the spectral data with~(\ref{C1}) or~(\ref{C2}). Taking into
account that $L_{0}$ is independent of $t$ by definition, we obtain
\begin{equation}
i\frac{\partial U}{\partial t}=[A,L]  \label{dUdt}
\end{equation}
for the potential $U$ from~(\ref{12}) and~(\ref{der:M}). Due to~(\ref{a11}),
the kernel of the l.h.s.\ is independent of ${\bf q}$ in the ($p,{\bf q}$)
representation, and $A(p;{\bf q})$ decays as ${\bf q}_{1}\rightarrow\infty$.
Taking~(\ref{12}) and~(\ref{13}) into account, we see that commutators with
$iD_{2}$ and $U$ in the r.h.s.\ also decay; the nonzero result therefore
gives only the commutator with $D_{1}^{2}$,
\[
i\frac{\partial U}{\partial t}=-\partial _{x_{1}}^{2}A-2(\partial
_{x_{1}}^{}A)D_{1}^{},
\]
where we introduce
\begin{equation}
\partial _{x_{j}}^{n}A=\underbrace{[D_{j}^{},\ldots ,[D_{j}^{}}_{n\,
{\rm times}},A]\ldots]  \label{commm}
\end{equation}
such that $(\partial_{x_{j}}A)(x,x',q)=A_{x_{j}}(x,x',q)+A_{x_{j}'}(x,x',q)$
in the $x$ representation. Again, $\partial_{x_{1}}^{2}A$ decays as
${\bf q}_{1}\rightarrow\infty$; therefore
\begin{equation}
\frac{\partial U}{\partial t}=2\partial _{x_{1}}^{}A_{}^{(-1)},
\label{dUdt1}
\end{equation}
where $A^{(-1)}$ is the operator whose kernel $A^{-1}(p,{\bf q})$ is equal to
the residue of $A(p;{\bf q})$ at infinity, i.e.,
\begin{equation}
A_{}^{(-1)}(p;{\bf q})=\lim_{{\bf q}_{1}\rightarrow \infty }A(p;{\bf q})
{\bf q}_{1}.  \label{res}
\end{equation}
From~(\ref{C1}), we have
\begin{equation}
A_{}^{(-1)}(p;{\bf q})=\frac{1}{2\pi }\int d{\bf q}'_{1\Re }\!\left(
|\nu _{}^{-}\rangle \frac{\partial F}{\partial t}\langle \omega
_{}^{+}|\right) \!(p;{\bf q}')  \label{res1}
\end{equation}
and from~(\ref{C2}),
\begin{equation}
A(x,x';q)=\delta (x_{1}^{}-x_{1}')\left( |\nu
_{}^{-}\rangle \frac{\partial F}{\partial t}\langle \omega
_{}^{+}|\right) \!(x,x';q)  \label{res2}
\end{equation}
in the $x$ representation. By construction, the last factor is proportional
to $\delta(x_{2}-x_{2}')$ and is independent of $q$. We can therefore see
that the r.h.s.\ of~(\ref{dUdt1}) is indeed a multiplication operator in the
$x$ representation, as it must be according to~(\ref{a11}).

\subsection{Linear symmetries of the spectral data}

To obtain integrable evolution equations associated with the NS equation from
this construction, we must consider linear symmetries of the spectral data
$F$. We therefore impose the condition that there exist operators $a^{\pm }$
such that
\begin{equation}
i\frac{\partial F}{\partial t}=a_{}^{-}F-Fa_{}^{+}.  \label{der:F}
\end{equation}
Because of~(\ref{sp}), we can chose kernels of these operators in the
($p,{\bf q}$) representation that are independent of ${\bf q}_{2}$ and
${\bf q}_{1\Im}$, i.e., without loss of generality,
\begin{equation}
a^{\pm }(p;{\bf q})=a^{\pm }(p;{\bf q}_{1\Re }).  \label{a:3}
\end{equation}
Properties~(\ref{F:2}) of the spectral data show that we can impose the
conditions
\begin{equation}
a_{}^{+}=(a^{-})_{}^{\dagger },\qquad [L_{0}^{},a_{}^{\pm }]=0,
\label{a1}
\end{equation}
again without loss of generality.

Returning to~(\ref{disc:A}) and using~(\ref{spJ}) and~(\ref{der:F}), we obtain
\begin{equation}
A_{}^{+}-|\nu _{}^{+}\rangle a_{}^{+}\langle \omega
_{}^{+}|=A_{}^{-}-|\nu _{}^{-}\rangle a_{}^{-}\langle \omega
_{}^{-}|.  \label{disc:A1}
\end{equation}
By construction, $A^{\pm}$, $|\nu^{\pm}\rangle$, and $\langle\omega^{\pm}|$
have analytic continuations in the upper and lower half-planes. It is
therefore natural to impose the condition that there exists a function
$a(p;{\bf q})$ (analytic or meromorphic) in ${\bf q}_{1}$ for
${\bf q}_{1\Im}\neq0$ and independent of ${\bf q}_{2}$ such that
\begin{equation}
a_{}^{\pm }(p;{\bf q})=\lim_{{\bf q}_{1\Im }\rightarrow \pm 0}a(p;{\bf q}).
\label{disc:a}
\end{equation}
We can thus consider different symmetries described by operators $a$ with
kernels $a(p;{\bf q})$ that have singularities of different types in the
complex plane. Here, we mainly consider the case where $a(p;{\bf q})$ depends
on ${\bf q}_{1}$ polynomially. Because, as is well known, the expansion
coefficients at infinity of the Jost solutions of the NS equation are given
by recursion relations in terms of the potential $u$, we call such symmetries
{\sl explicit.} In particular, because $a$ has no discontinuity on the real
axis for such symmetries, i.e.,
\begin{equation}
a^{-}=a^{+},  \label{a:4}
\end{equation}
conditions~(\ref{a1}) give
\begin{equation}
a=a_{}^{\dagger },\qquad [L_{0}^{},a]=0.  \label{a5}
\end{equation}
We then have
\begin{equation}
i\frac{\partial F}{\partial t}=[a,F]  \label{der:F1}
\end{equation}
instead of~(\ref{der:F}), and~(\ref{C1}) can be written as
\begin{eqnarray}
&&A=(|\nu \rangle a\langle \omega |)_{-}^{},  \label{bilin:A} \\
&&P=(|\nu \rangle a\langle \omega |)_{+}^{},  \label{bilin:P} \\
&&|\nu \rangle a\langle \omega |=A+P,  \label{AP}
\end{eqnarray}
where $(\cdot)_{+}$ and $(\cdot)_{-}$ denote positive (polynomial) and
negative parts of the $1/{\bf q}_{1}$ expansion at infinity of the kernels in
the ($p,{\bf q}$) representation. From relations~(\ref{bilin:A})--(\ref{AP})
with~(\ref{2326}) and the commutativity of $a$ with $L_{0}$ and with $M_{0}$
taken into account, we obtain
\begin{eqnarray}
&&i\frac{\partial M}{\partial t}=[P,M],\qquad\qquad\,\,
i\frac{\partial L}{\partial t}=[P,L],  \label{der1} \\
&&i\frac{\partial |\nu \rangle }{\partial t}=P|\nu \rangle -|\nu \rangle
a,\qquad i\frac{\partial \langle \omega |}{\partial t}=a\langle \omega
|-\langle \omega |P  \label{der2}
\end{eqnarray}
instead of~(\ref{der:M}) and~(\ref{der:J}).

It is easy to prove that all polynomials $a$ with properties~(\ref{a5}) can
be obtained as linear combinations of the operators
\begin{equation}
a_{m,n}^{}=\frac{i_{}^{m}}{2}\{D_{1}^{m},\Delta _{}^{n}\},\quad
m,n\geq 0,\quad m+n>0.  \label{amn}
\end{equation}
Here,
\begin{equation}
\Delta =X_{1}^{}+2iX_{2}^{}D_{1}^{},  \label{D'}
\end{equation}
and we introduce the operators $X_{j}$ that are operators of multiplication
by $x_{j}$ in the $x$ representation,
\begin{equation}
X_{j}^{}(x,x';q)=x_{j}^{}\delta (x-x'),\quad j=1,2.
\label{def:Xj}
\end{equation}
It is easy to verify that
\begin{eqnarray}
&&X_{j}^{\dag }=X_{j}^{},\qquad [D_{j}^{},X_{k}^{}]=\delta _{j,k}^{},
\label{X} \\
&&\Delta _{}^{\dag }=\Delta ,\qquad\,\,\, [D_{1}^{},\Delta ]=I,\qquad
[D_{2}^{},\Delta ]=2iD_{1}^{},  \label{Delta} \\
&&[L_{0}^{},\Delta ]=0.  \label{L0D}
\end{eqnarray}
Symmetries generated by the operators $a_{m,n}$ were called {\sl additional}
in~\cite{OrlovS}. We here consider time evolutions associated with their
subclasses.

\subsection{Subsets of commuting symmetries and hierarchies of evolutions}

In general, the flows introduced in~(\ref{amn}) do not commute. On the other
hand, it is clear that in the linear span of these operators, there exist
subsets of mutually commuting operators. We consider two operators $a$ and
$a'$, let $t$ and $t'$ be the corresponding evolution parameters of these
symmetries, and let $A$, $A'$ and $P$, $P'$ be defined  by $a$ and $a'$ as in
(\ref{bilin:A}) and~(\ref{bilin:P}). In terms of the spectral data, the
commutativity condition for these flows is simply
\begin{equation}
\lbrack a,a']=0,  \label{acomm}
\end{equation}
where we assume that $a$ and $a'$ are independent of $t$ and $t'$. In terms
of $A$ and $P$, this condition is more involved. Indeed, from~(\ref{der2}),
\[
\frac{\partial }{\partial t}\Bigl(|\nu \rangle a'\langle \omega |
\Bigr)=i\Bigl[A,|\nu \rangle a'\langle \omega |\Bigr]
\]
and vice versa; we therefore have
\begin{equation}
\frac{\partial }{\partial t}\Bigl(|\nu \rangle a'\langle \omega |
\Bigr)-\frac{\partial }{\partial t'}\Bigl(|\nu \rangle a\langle
\omega |\Bigr)=i|\nu \rangle [a,a']\langle \omega |-i[A',A]+i[P',P].
\label{ZS'}
\end{equation}
In case~(\ref{acomm}), the polynomial and the decreasing parts of this
equality can be separated, and we obtain the Zakharov--Shabat conditions
\begin{equation}
\frac{\partial A}{\partial t'}-\frac{\partial A'}{\partial
t}=i[A,A'],\qquad \frac{\partial P}{\partial t'}-
\frac{\partial P'}{\partial t}=i[P,P'].  \label{ZS"}
\end{equation}

As we see from~(\ref{amn}) and~(\ref{D'}), $a_{m,n}$ is a differential
operator of order $m+n$. If $m$ and $n$ are not relatively prime, i.e., if
there exist numbers $k\ne1$, $m'$, and $n'$ such that $m=m'k$ and $n=n'k$,
then $a_{m,n}$ is equal to $(a_{m',n'})^{k}$ plus a linear combination of
some lower $a_{m'',n''}$, $m''+n''<m+n$. The linear span of set (\ref{amn})
is thus decomposed into the union of subsets of commuting flows. These
subsets are labeled by pairs of relatively prime numbers ($k,l$), the
operator of the lowest order in the ($k,l$) subset is $a_{k,l}$, and all
elements of this subset are equal to $(a_{k,l})^{m}$, $m=1,2,\ldots$ (or to
their linear combinations). Because these subsets of mutually commuting
operators are disjoint, we can consider only symmetries belonging to one
($k,l$) subset. For simplicity, let $a$ denote $a_{k,l}$. We thus study
symmetries determined by the operators $a^{m}$, $m=1,2,\ldots $, and let the
corresponding evolution parameters be denoted by $t_{m}$,
\begin{equation}
i\frac{\partial F}{\partial t_{m}^{}}=[a_{}^{m},F],\quad m=1,2,\ldots\,.
\label{der:Fm}
\end{equation}
To emphasize the commutativity of these symmetries, we use the term {\sl
times} for these parameters.

Using~(\ref{der:Fm}) and inverse problem~(\ref{spJ}) for the Jost solutions,
we can introduce dependence on these times for all objects of the theory;
thus, by~(\ref{der1}) and~(\ref{der2}),
\begin{eqnarray}
&&i\frac{\partial M}{\partial t_{m}^{}}=[P_{m}^{},M],\qquad\qquad\quad
i\frac{\partial L}{\partial t_{m}^{}}=[P_{m}^{},L],  \label{der1m} \\
&&i\frac{\partial |\nu \rangle }{\partial t_{m}^{}}=P_{m}^{}|\nu \rangle
-|\nu \rangle a^{m}_{},\qquad i\frac{\partial \langle \omega |}{\partial
t_{m}^{}}=a_{}^{m}\langle \omega |-\langle \omega |P_{m}^{},
\label{der2m}
\end{eqnarray}
where by~(\ref{bilin:P}),
\begin{equation}
P_{m}^{}=(|\nu \rangle a_{}^{m}\langle \omega |)_{+}^{}.  \label{Pm}
\end{equation}
To be more precise, we mark the original operators with a hat: $\widehat{F}$,
$\widehat{U}$, $|\hat{\nu}\rangle$, $\langle\hat{\omega}|$, etc. Then the
$t$-dependent operators $F(t_{1},t_{2},\ldots )$, etc., are determined by the
corresponding initial data,
\begin{equation}
F(t_{1}^{},t_{2}^{},\ldots )|_{t_{1}=t_{2}=\ldots =0}=\widehat{F},
\label{Cauchy}
\end{equation}
etc. We emphasize that all time variables are here introduced in addition to
the original variables of the kernels of all operators participating in the
construction of the spectral theory of the NS equation. It is clear that the
time variables are distinct from the variables of the kernels of operators;
in particular, the derivatives with respect to these sets of variables
commute.

The times $t_{m}$ are naturally ordered by powers of the differential
operators $a^{m}$. We mention that evolutions~(\ref{der:Fm}) can also be
considered if  $a(p;{\bf q})$ is a meromorphic function in the complex
domain of ${\bf q}_{1}$. In this case, of course, $P_{m}$ stands for the
singular part of the Laurent expansion of $|\nu\rangle a^{m}\langle\omega|$;
it is then given in terms of not the potential $U$ but the values of the Jost
solutions at the poles of $a$. It is important that in all these cases {\sl
independently of the exact choice of operator} $a$, evolutions~(\ref{der:Fm})
are associated with the KP equation and its hierarchy. Indeed, for arbitrary
operators $a$ and $F$, there exists the simple commutator identity
\begin{equation}
\lbrack a_{}^{3},[a,F]]-\frac{3}{4}[a_{}^{2},[a_{}^{2},F]]+\frac{1}{4}
[a,[a,[a,[a,F]]]]=0,  \label{commutator}
\end{equation}
which implies that the spectral data $F$ satisfy the differential equation
\begin{equation}
\frac{\partial _{}^{2}F}{\partial t_{1}^{}\partial t_{3}^{}}-\frac{3}{4}
\frac{\partial _{}^{2}F}{\partial t_{2}^{2}}+\frac{1}{4}\frac{\partial
_{}^{4}F}{\partial t_{1}^{4}}=0,  \label{lKPI}
\end{equation}
i.e., the linearized KPI equation, and so on for higher linearized equations.

Moreover, because of~(\ref{der2m}), we have the operator equations
\begin{eqnarray}
&&i\frac{\partial |\nu \rangle }{\partial t_{1}^{}}=P_{1}^{}|\nu \rangle
-|\nu \rangle a,  \label{op} \\
&&i\frac{\partial |\nu \rangle }{\partial t_{2}^{}}+\frac{\partial
_{}^{2}|\nu \rangle }{\partial t_{1}^{2}}=\left( P_{2}^{}-P_{1}^{2}-i
\frac{\partial P_{1}^{}}{\partial t_{1}^{}}\right) |\nu \rangle +2i
\frac{\partial |\nu \rangle }{\partial t_{1}^{}}a,  \nonumber \\
&&i\frac{\partial|\nu\rangle}{\partial t_{3}^{}}+i\frac{\partial
^{3}|\nu\rangle}{\partial t_{1}^{3}}=\left( P_{3}^{}+\frac{\partial
_{}^{2}P_{1}^{}}{\partial t_{1}^{2}}-2i\frac{\partial P_{1}^{}}
{\partial t_{1}^{}}P_{1}^{}-iP_{1}^{}\frac{\partial P_{1}^{}}
{\partial t_{1}^{}}-P_{1}^{3}\right) |\nu \rangle +  \nonumber \\
&&\phantom{i\frac{\partial|\nu\rangle}{\partial t_{3}^{}}+i\frac{\partial
^{3}|\nu\rangle}{\partial t_{1}^{3}}={}}
+3i\frac{\partial (P_{1}^{}|\nu \rangle )}{\partial t_{1}^{}}a,
\nonumber
\end{eqnarray}
which, in the case where $\partial P_{1}/\partial t_{1}=0$, are simplified to
\begin{eqnarray}
&&i\frac{\partial |\nu \rangle }{\partial t_{2}^{}}+\frac{\partial
_{}^{2}|\nu \rangle }{\partial t_{1}^{2}}=\left(
P_{2}^{}-P_{1}^{2}\right) |\nu \rangle +2i\frac{\partial |\nu \rangle }
{\partial t_{1}^{}}a,  \label{op1} \\
&&i\frac{\partial |\nu \rangle }{\partial t_{3}^{}}+i\frac{\partial
^{3}|\nu \rangle }{\partial t_{1}^{3}}=\left( P_{3}^{}-P_{1}^{3}\right)
|\nu \rangle +3iP_{1}^{}\frac{\partial |\nu \rangle }{\partial t_{1}^{}}a;
\label{op2}
\end{eqnarray}
here, the first equation is the obvious analogue of~(\ref{2011}) with the
operators $P_{2}^{}-P_{1}^{2}$ and $a$ playing the roles of the potential and
of the spectral parameter. Using~(\ref{op}) and~(\ref{op1}), we can rewrite
(\ref{op2}) as
\begin{eqnarray*}
&&\frac{\partial|\nu\rangle}{\partial t_{3}^{}}-\frac{3i}{2}
\frac{\partial^{2}|\nu\rangle}{\partial t_{1}^{}\partial t_{2}^{}}-
\frac{1}{2}\frac{\partial^{3}|\nu\rangle}{\partial t_{1}^{3}}=
-i\left( P_{3}^{}+ \frac{1}{2}P_{1}^{3}-
\frac{3}{2}P_{1}^{}P_{2}^{}\right) |\nu \rangle - \\
&&\phantom{\frac{\partial|\nu\rangle}{\partial t_{3}^{}}-\frac{3i}{2}
\frac{\partial^{2}|\nu\rangle}{\partial t_{1}^{}\partial t_{2}^{}}-
\frac{1}{2}\frac{\partial^{3}|\nu\rangle}{\partial t_{1}^{3}}={}}
-\frac{3i}{2}P_{1}^{}\left( i\frac{\partial |\nu \rangle }
{\partial t_{2}^{}}+\frac{\partial _{}^{2}|\nu \rangle }{\partial
t_{1}^{2}}\right) a.
\end{eqnarray*}
The derivative of this equation with respect to $t_{1}$ after $\frac{\partial
|\nu\rangle}{\partial t_{1}}a$ is replaced using~(\ref{op1}) gives the
``dressing'' of Eq.~(\ref{lKPI})
\begin{eqnarray}
&&\frac{\partial_{}^{2}|\nu\rangle}{\partial t_{1}^{}\partial
t_{3}^{}}-\frac{3}{4}\frac{\partial^{2}|\nu\rangle}{\partial t_{2}^{2}}+
\frac{1}{4}\frac{\partial^{4}|\nu\rangle}{\partial t_{1}^{4}}=
\left( \left[ P_{3}^{}+\frac{1}{2}P_{1}^{3}-\frac{3}{4}
\{P_{1}^{},P_{2}^{}\},P_{1}^{}\right] +\frac{3i}{4}\frac{\partial
P_{2}^{}}{\partial t_{2}^{}}\right) |\nu \rangle -  \nonumber \\
&&\phantom{\frac{\partial_{}^{2}|\nu\rangle}{\partial t_{1}^{}\partial
t_{3}^{}}-\frac{3}{4}\frac{\partial^{2}|\nu\rangle}{\partial t_{2}^{2}}+
\frac{1}{4}\frac{\partial^{4}|\nu\rangle}{\partial t_{1}^{4}}={}}
-i\left( P_{3}^{}+\frac{1}{2}P_{1}^{3}-\frac{3}{2}
P_{2}^{}P_{1}^{}\right) \frac{\partial |\nu \rangle }{\partial t_{1}^{}
}+  \nonumber \\
&&\phantom{\frac{\partial_{}^{2}|\nu\rangle}{\partial t_{1}^{}\partial
t_{3}^{}}-\frac{3}{4}\frac{\partial^{2}|\nu\rangle}{\partial t_{2}^{2}}+
\frac{1}{4}\frac{\partial^{4}|\nu\rangle}{\partial t_{1}^{4}}={}}
+\frac{3}{4}(P_{2}^{}-P_{1}^{2})\left( i\frac{\partial |\nu \rangle }
{\partial t_{2}^{}}+\frac{\partial _{}^{2}|\nu \rangle }{\partial
t_{1}^{2}}\right) ,  \label{op3}
\end{eqnarray}
where we use~(\ref{ZS"}) and $\{\cdot,\cdot\}$ stands for the anticommutator
of operators.

The simplest examples of polynomial flows of type~(\ref{der:Fm}) are given by
two special subsets (1,0) and (0,1), which, by~(\ref{amn}), are respectively
\begin{eqnarray}
&&a_{1,0}=iD_{1}^{},  \label{flows1} \\
&&a_{0,1}=\Delta .  \label{flows2}
\end{eqnarray}
We consider these evolutions in more detail below.

\section{Examples of the simplest evolutions}

\subsection{Times $t_{m,0}$}

We first briefly demonstrate the above technique applied to the standard
flows~(\ref{flows1}). The corresponding times and polynomials are denoted by
$t_{m,0}$ and $P_{m,0}$. We thus have
\begin{eqnarray}
&&i\frac{\partial F}{\partial t_{m,0}^{}}=[(iD_{1}^{})_{}^{m},F],
\quad m=1,2,\ldots ,  \label{der:Fm0} \\
&&i\frac{\partial |\nu \rangle }{\partial t_{m,0}^{}}=P_{m,0}^{}|\nu
\rangle -|\nu \rangle (iD_{1}^{})_{}^{m},  \label{der:num0} \\
&&P_{m,0}^{}=(|\nu \rangle (iD_{1}^{})_{}^{m}\langle \omega
|)_{+}^{}.  \label{Pm0}
\end{eqnarray}

Using representation~(\ref{sp}) for the kernel of operator $F$, we obtain
time evolutions of the spectral data in the standard form
\begin{equation}
\frac{\partial f(\alpha ,\beta )\ }{\partial t_{m,0}}=i[(\alpha
_{}^{m}-\beta _{}^{m}]f(\alpha ,\beta ).  \label{der:f1}
\end{equation}
To calculate the polynomials $P_{m,0}$ explicitly, we use the asymptotic
expansions of the Jost solutions
\begin{equation}
|\nu \rangle =\sum_{n=0}^{\infty }\nu _{n}(2iD_{1})^{-n},\qquad \langle
\omega |=\sum_{n=0}^{\infty }(2iD_{1})^{-n}\omega _{n},  \label{exp:J}
\end{equation}
which have the exact meaning in our approach of the $1/{\bf q}_{1}$ expansion
at infinity for $|\nu\rangle(p;{\bf q})$ and the $1/({\bf q}_{1}+p_{1})$
expansion for $\langle\omega|(p;{\bf q})$. It is known that the coefficients
of these expansions depend on the half-plane of ${\bf q}_{1}$, but here we
can ignore this fact, referring to~\cite{kpshort} and~\cite{KPJMP} for
details. To calculate these coefficients, we can use either the integral
equation for the Jost solutions or bilinear relations~(\ref{2326}) and the
relation
\begin{equation}
U=iD_{2}^{}+D_{1}^{2}-|\nu \rangle (iD_{2}^{}+D_{1}^{2})\langle \omega |,
\label{Udress}
\end{equation}
which follows from~(\ref{12}),~(\ref{13}), and~(\ref{24}). We then obtain
\begin{eqnarray}
&&\nu _{0}=\omega _{0}=I,  \label{exp:0} \\
&&\nu _{1}=-\omega _{1},\qquad \partial _{x_{1}}^{}\nu _{1}^{}=iU,
\label{exp:1} \\
&&\omega _{2}^{}+\nu _{2}^{}=-\nu _{1}^{}\omega _{1}-i\partial
_{x_{1}}^{}(\nu _{1}^{}-\omega _{1}),  \label{exp:2} \\
&&\partial _{x_{1}}^{}\nu _{2}^{}=\partial _{x_{2}}^{}\nu
_{1}^{}-i\partial _{x_{1}}^{2}\nu _{1}^{}+\frac{1}{2}\partial
_{x_{1}}^{}\nu _{1}^{2},  \label{exp:2'}
\end{eqnarray}
and so on, where notation~(\ref{commm}) is used. We omit the corresponding
recursion relations because we do not need the highest coefficients here.
Using these formulas and~(\ref{Pm0}), we obtain the combinations involved in
(\ref{op}),~(\ref{op1}), and~(\ref{op3}):
\begin{eqnarray}
&&P_{1,0}^{}=iD_{1}^{},  \label{10:P} \\
&&P_{2,0}^{}-P_{1,0}^{2}=U,  \label{20:P1} \\
&&P_{3,0}^{}+\frac{1}{2}P_{1,0}^{3}-\frac{3}{2}P_{2,0}^{}P_{1,0}^{}=
\frac{3i}{4}(\partial _{x_{1}}^{}U-\partial _{x_{2}}^{}\nu _{1}^{}).
\label{30:P}
\end{eqnarray}
For the first expression in~(\ref{op3}), we have
\begin{equation}
\left[ P_{3,0}^{}+\frac{1}{2}P_{1,0}^{3}-\frac{3}{4}
\{P_{2,0}^{},P_{1,0}^{}\},P_{1,0}^{}\right] +\frac{3i}{4}
\frac{\partial P_{2,0}^{}}{\partial t_{2,0}^{}}=\frac{3i}{4}\left(
\frac{\partial U}{\partial t_{2,0}^{}}-\partial _{x_{2}}^{}U\right) \!.
\label{30:P1}
\end{equation}
By~(\ref{op}) and~(\ref{op1}), $|\nu\rangle$ thus satisfies the equations
\begin{eqnarray}
&&\frac{\partial |\nu \rangle }{\partial t_{1,0}^{}}=\partial
_{x_{1}}^{}|\nu \rangle ,  \label{10:J} \\
&&i\frac{\partial |\nu \rangle }{\partial t_{2,0}^{}}+\frac{\partial
_{}^{2}|\nu \rangle }{\partial t_{2,0}^{2}}=U|\nu \rangle -2\frac{\partial
|\nu \rangle }{\partial t_{1,0}^{}}D_{1}^{},  \label{20:J}
\end{eqnarray}
which, due to~(\ref{12}),~(\ref{13}),~(\ref{23}),~(\ref{commm}), and~(\ref
{10:J}), are exactly
\begin{equation}
\frac{\partial |\nu \rangle }{\partial t_{2,0}^{}}=\partial
_{x_{2}}^{}|\nu \rangle .  \label{20:nu}
\end{equation}
Correspondingly, from~(\ref{exp:J}) and~(\ref{exp:1}), we obtain the
equalities for the potential $U$
\begin{equation}
\frac{\partial U}{\partial t_{1,0}^{}}=\partial _{x_{1}}^{}U,\qquad
\frac{\partial U}{\partial t_{2,0}^{}}=\partial _{x_{2}}^{}U.
\label{1020:U}
\end{equation}
Using representations~(\ref{a11}) and~(\ref{23c}), we obtain
\begin{eqnarray}
&&\frac{\partial \chi (x,k)}{\partial t_{1,0}^{}}=\chi
_{x_{1}}^{}(x,k),\qquad \frac{\partial \chi (x,k)}{\partial t_{2,0}^{}}
=\chi _{x_{2}}^{}(x,k),  \label{1020:chi} \\
&&\frac{\partial u(x)}{\partial t_{1,0}^{}}=u_{x_{1}}^{}(x),\qquad\qquad
\frac{\partial u(x)}{\partial t_{2,0}^{}}=u_{x_{2}}^{}(x).  \label{1020:u}
\end{eqnarray}
These equalities are often considered the reason to identify $t_{1,0}=x_{1}$
and $t_{2,0}=x_{2}$. Here, as mentioned in discussing
(\ref{der1m})--(\ref{Cauchy}), we write
\begin{eqnarray}
&&\chi (t_{1,0}^{},t_{2,0}^{}|x_{1}^{},x_{2}^{},k)=\widehat{\chi }
(x_{1}^{}+t_{1,0}^{},x_{2}^{}+t_{2,0}^{},k),  \nonumber \\
&&u(t_{1,0}^{},t_{2,0}^{}|x_{1}^{},x_{2}^{})=\widehat{u}
(x_{1}^{}+t_{1,0}^{},x_{2}^{}+t_{2,0}^{}).  \label{shiftx}
\end{eqnarray}

Now, we see that due to~(\ref{1020:U}), expression~(\ref{30:P1}) is equal to
zero; (\ref{op3}) is therefore reduced to
\begin{eqnarray}
\frac{\partial _{}^{2}|\nu \rangle }{\partial t_{1,0}^{}\partial
t_{3,0}^{}}-\frac{3}{4}\frac{\partial ^{2}|\nu \rangle }{\partial
t_{2,0}^{2}}+\frac{1}{4}\frac{\partial ^{4}|\nu \rangle }{\partial
t_{1,0}^{4}}=&&\frac{3}{4}\left( \partial _{x_{1}}^{}U-\partial
_{x_{2}}^{}\nu _{1}^{}\right) \frac{\partial |\nu \rangle }{\partial
t_{1,0}^{}}+  \nonumber \\
&&+\frac{3U}{4}\left( i\frac{\partial |\nu \rangle }{\partial
t_{2,0}^{}}+\frac{\partial _{}^{2}|\nu \rangle }{\partial t_{1,0}^{2}}
\right) \!;  \label{30:J}
\end{eqnarray}
and from~(\ref{exp:J}) and~(\ref{exp:1}), we obtain the standard KPI equation
\begin{equation}
\frac{\partial _{}^{2}U}{\partial t_{1,0}^{}\partial t_{3,0}^{}}-
\frac{3}{4}\frac{\partial ^{2}U}{\partial t_{2,0}^{2}}+\frac{1}{4}\frac
{\partial ^{4}U}{\partial t_{1,0}^{4}}=\frac{3}{4}\frac{\partial _{}^{2}
U_{}^{2}}{\partial t_{1,0}^{2}}.  \label{KPI}
\end{equation}

\subsection{Times $t_{0,m}$}

In this section, we investigate evolutions determined by~(\ref{flows2}). The
corresponding times and polynomials are denoted by $t_{0,m}$ and $P_{0,m}$,
and we again emphasize that derivatives with respect to $t_{0,m}$ commute
with the $x$ variables and the times $t_{m,0}$ considered above are not
switched on now. Here, we thus have
\begin{eqnarray}
&&i\frac{\partial F}{\partial t_{0,m}^{}}=[\Delta _{}^{m},F],\quad
m=1,2,\ldots ,  \label{der:F0m} \\
&&i\frac{\partial |\nu \rangle }{\partial t_{0,m}^{}}=P_{0,m}^{}|\nu
\rangle -|\nu \rangle \Delta _{}^{m},  \label{der:nu0m} \\
&&P_{0,m}^{}=(|\nu \rangle \Delta _{}^{m}\langle \omega |)_{+}^{},
\label{P0m}
\end{eqnarray}
where the operator $\Delta$ is defined in~(\ref{D'}). Time evolutions of the
standard spectral data~(\ref{sp}) are now
\begin{equation}
\ \frac{\partial f(\alpha ,\beta )\ }{\partial t_{0,n}}=i\left[ \left( -i
\frac{\partial }{\partial \alpha }\right) ^{n}-\left( i\frac{\partial }
{\partial \beta }\ \right) ^{n}\right] f(\alpha ,\beta ),   \label{der:f2}
\end{equation}
and for simplicity in what follows, we consider only $m=1,2,3$.

The polynomials $P_{0,m}$ can be explicitly written using (\ref{exp:J}),
(\ref{exp:0}), and~(\ref{exp:2'}); it must be taken into account that $X_{2}$
commutes with $|\nu\rangle$ and $\langle\omega|$ because of~(\ref{181}),
(\ref{182}), and~(\ref{def:Xj}). For the combinations involved in~(\ref{op}),
(\ref{op1}), and~(\ref{op3}), we thus obtain
\begin{eqnarray}
&&P_{0,1}^{}=\Delta ,  \label{01:P} \\
&&P_{0,2}^{}-P_{0,1}^{2}=4X_{2}^{2}U,  \label{02:P} \\
&&P_{0,3}^{}+\frac{1}{2}P_{0,1}^{3}-\frac{3}{2}P_{0,2}^{}P_{0,1}^{}=
6iX_{2}^{3}(\partial _{x_{1}}^{}U-\partial _{x_{2}}^{}\nu
_{1}^{})-6iX_{2}^{2}\partial _{x_{1}}^{}(X_{1}^{}\nu _{1}^{}),
\label{03:P}
\end{eqnarray}
and for the first expression in~(\ref{op3}), we have
\begin{eqnarray}
&&\left[ P_{0,3}^{}+\frac{1}{2}P_{0,1}^{3}-\frac{3}{4}
\{P_{0,2}^{},P_{0,1}^{}\},P_{0,1}^{}\right] +\frac{3i}{4}
\frac{\partial P_{0,2}^{}}{\partial t_{0,2}^{}}=  \nonumber \\
&&\qquad\qquad\qquad\qquad\qquad
=3iX_{2}^{2}\left( \frac{\partial U}{\partial t_{0,2}^{}}
-4X_{2}^{}[\partial _{x_{1}}^{}(X_{1}^{}U\!)+\partial
_{x_{2}}^{}(X_{2}^{}U\!)]\right).  \label{03:P1}
\end{eqnarray}
Then, by~(\ref{op}) and~(\ref{op1}),
\begin{eqnarray}
&&\frac{\partial |\nu \rangle }{\partial t_{0,1}^{}}=-i[\Delta ,|\nu
\rangle ]=-i[X_{1}^{},|\nu \rangle ]+2X_{2}^{}\partial _{x_{1}}^{}|\nu
\rangle ,  \label{01:nu} \\
&&i\frac{\partial |\nu \rangle }{\partial t_{0,2}^{}}+\frac{\partial
^{2}|\nu \rangle }{\partial t_{0,1}^{2}}=2i\frac{\partial |\nu \rangle }
{\partial t_{0,1}^{}}\Delta +4X_{2}^{2}U|\nu\rangle,  \label{02:nu}
\end{eqnarray}
and the second equation, i.e., the NS equation with respect to these times,
is thus the exact analogue of~(\ref{20:J}) with the role of the potential
played by the multiplication operator (cf.~(\ref{a11}))
\begin{equation}
W=4X_{2}^{2}U,\qquad W(x,x',q)=w(x)\delta (x-x'),\qquad
w(x)=4x_{2}^{2}u(x).  \label{W}
\end{equation}

Expanding these equations by~(\ref{exp:J}) and using~(\ref{exp:0}) and
(\ref{exp:1}), we derive
\begin{equation}
\frac{\partial U}{\partial t_{0,1}^{}}=2X_{2}^{}\partial
_{x_{1}}^{}U,\qquad \frac{\partial U}{\partial t_{0,2}^{}}
=4X_{2}^{}[\partial _{x_{1}}^{}(X_{1}^{}U)+\partial
_{x_{2}}^{}(X_{2}^{}U)],  \label{0102:U}
\end{equation}
where we take $[X_{1},|\nu\rangle]\sim D_{1}^{-2}$ into account.

The evolution with respect to $t_{0,3}$ via~(\ref{op3}) is again simplified
because the expression in~(\ref{03:P1}) is equal to zero due to
(\ref{0102:U}). From~(\ref{02:P}) and~(\ref{03:P}), we then obtain
\begin{eqnarray}
&&\frac{\partial_{}^{2}|\nu\rangle}{\partial t_{0,1}^{}\partial
t_{0,3}^{}}-\frac{3}{4}\frac{\partial^{2}|\nu\rangle}{\partial
t_{0,2}^{2}}+\frac{1}{4}\frac{\partial^{4}|\nu\rangle}{\partial
t_{0,1}^{4}}=6X_{2}^{2}\left( X_{2}^{}(\partial _{x_{1}}^{}U-\partial
_{x_{2}}^{}\nu _{1}^{})-\partial _{x_{1}}^{}(X_{1}^{}\nu
_{1}^{})\right) \frac{\partial |\nu \rangle }{\partial t_{0,1}^{}}
+\qquad   \nonumber \\
&&\phantom{\frac{\partial_{}^{2}|\nu\rangle}{\partial t_{0,1}^{}\partial
t_{0,3}^{}}-\frac{3}{4}\frac{\partial^{2}|\nu\rangle}{\partial
t_{0,2}^{2}}+\frac{1}{4}\frac{\partial^{4}|\nu\rangle}{\partial
t_{0,1}^{4}}={}}
+3X_{2}^{2}U\left( i\frac{\partial |\nu \rangle }{\partial
t_{0,2}^{}}+\frac{\partial _{}^{2}|\nu \rangle }{\partial t_{0,1}^{2}}
\right) \!.  \label{03:nu}
\end{eqnarray}
Using the above relations, we can extract the term of order $D_{1}^{-1}$.
For the potential $W$ in Eq.~(\ref{02:nu}), we then obtain exactly the KPI
equation in terms of the times $t_{0,1}$, $t_{0,2}$, and $t_{0,3}$:
\begin{equation}
\frac{\partial _{}^{2}W}{\partial t_{0,1}^{}\partial t_{0,3}^{}}-
\frac{3}{4}\frac{\partial ^{2}W}{\partial t_{0,2}^{2}}+\frac{1}{4}
\frac{\partial ^{4}W}{\partial t_{0,1}^{4}}=\frac{3}{4}
\frac{\partial _{}^{2}W_{}^{2}}{\partial t_{0,1}^{2}}.  \label{KPIW}
\end{equation}

\subsection{Transformation of dependent and independent variables}

As shown in the previous section, taking the time $t_{0,m}$ into account
gives rise to a new dependent variable $w(x)$ via~(\ref{W}). Because of
(\ref{KPIW}), this function of five variables, $w(t_{0,1},t_{0,2},t_{03}|x)$,
satisfies the KPI equation with respect to the $t_{0,m}$ variables,
\begin{equation}
\frac{\partial _{}^{2}w}{\partial t_{0,1}^{}\partial t_{0,3}^{}}-
\frac{3}{4}\frac{\partial ^{2}w}{\partial t_{0,2}^{2}}+\frac{1}{4}
\frac{\partial^{4}w}{\partial t_{0,1}^{4}}=\frac{3}{4}
\frac{\partial _{}^{2}w_{}^{2}}{\partial t_{0,1}^{2}},  \label{KPIw}
\end{equation}
for all $x$, and Eq.~(\ref{0102:U}), together with~(\ref{a11}) and~(\ref{W}),
implies
\begin{equation}
\frac{\partial w(x)}{\partial t_{0,1}^{}}=2x_{2}^{}w{}_{x_{1}}^{}(x),
\qquad \frac{\partial w(x)}{\partial t_{0,2}^{}}
=4x_{2}^{}[x_{1}^{}w(x)_{x_{1}}^{}+x_{2}^{}w(x)_{x_{2}}^{}],
\label{0102:w}
\end{equation}
which suggests a transformation of the independent variables as well. Indeed,
let
\begin{equation}
z_{1}^{}=\frac{x_{1}^{}}{2x_{2}^{}},\qquad
z_{2}^{}=\frac{-1}{4x_{2}^{}};  \label{z}
\end{equation}
then instead of~(\ref{0102:w}), we obtain
\begin{equation}
\frac{\partial w(x)}{\partial t_{0,1}^{}}=w{}_{z_{1}}^{}(x),\qquad
\frac{\partial w(x)}{\partial t_{0,2}^{}}=w{}_{z_{2}}^{}(x).  \label{0102}
\end{equation}
This means that $t_{0,1}$ and $t_{0,2}$ shift the variables $z_{1}$ and
$z_{2}$. In other words, if we introduce
\begin{equation}
\varphi (t_{0,1}^{},t_{0,2}^{}|z)=w(t_{0,1}^{},t_{0,2}^{}|x),
\label{wchi}
\end{equation}
then (cf.~(\ref{Cauchy}))
\[
\varphi (t_{0,1}^{},t_{0,2}^{}|z)=\widehat{\varphi }
(z_{1}^{}+t_{0,1}^{},z_{2}^{}+t_{0,2}^{});
\]
letting $\widehat{w}$ denote the value of $w$ at $t_{0,1}=t_{0,2}=0$, we
obtain
\begin{equation}
\widehat{w}(x)=\widehat{\varphi }\left( \frac{x_{1}^{}}{2x_{2}^{}},
\frac{-1}{4x_{2}^{}}\right)   \label{wphihat}
\end{equation}
and
\begin{equation}
w(t_{0,1}^{},t_{0,2}^{}|x)=\widehat{w}\left( \frac{x_{1}^{}+2x_{2}^{}
t_{0,1}^{}}{1-4x_{2}^{}t_{0,1}^{}},
\frac{x_{2}^{}}{1-4x_{2}^{}t_{0,1}^{}}\right) .  \label{whatw}
\end{equation}
We see that the class of initial data and, correspondingly, the class of
solutions of the KPI equation can be essentially extended in this way. Indeed,
we seek a solution $\varphi (t_{0,1},t_{0,2},t_{0,3})$ of Eq.~(\ref{KPIw})
that satisfies the initial data
\begin{equation}
\varphi (t_{0,1},t_{0,2},0)=\widehat{\varphi }(t_{0,1}^{},t_{0,2}^{}).
\label{phi0}
\end{equation}
Let
\begin{equation}
w(t_{0,1}^{},t_{0,2}^{},t_{0,3}^{}|x)=\varphi
(t_{0,1}^{}+z_{1}^{},t_{0,2}^{}+z_{2}^{},t_{0,3}),  \label{wphi3}
\end{equation}
where $z$ is given by~(\ref{z}). This function also solves~(\ref{KPIw}), and
by~(\ref{phi0}), Eq.~(\ref{wphihat}) gives $\widehat{w}(x)=w(0,0,0|x)$.
Switching on the times $t_{0,m}$ and constructing the function $u(t_{0,1},
t_{0,2},t_{0,3}|x)$ with the above procedure, we conclude from~(\ref{W}) and
(\ref{wphihat}) that it satisfies
\begin{equation}
\widehat{u}(x)=\frac{1}{4x_{2}^{2}}\widehat{\varphi }\left( \frac{x_{1}^{}}
{2x_{2}^{}},\frac{-1}{4x_{2}^{}}\right) .  \label{uphihat}
\end{equation}
Transforming the function $w(t_{0,1},t_{0,2},t_{0,3}|x)=4x_{2}^{2}u(t_{0,1},
t_{0,2},t_{0,3}|x)$ with the transformation inverse to~(\ref{z}) and setting
$z=0$, we obtain the solution of the Cauchy problem for $\varphi(t_{0,1},
t_{0,2},t_{0,3})$ from~(\ref{phi0}). By~(\ref{uphihat}), the equation is thus
solved for the initial data
\[
\widehat{\varphi }(t_{0,1}^{},t_{0,2}^{})=\frac{1}{4t_{0,2}^{2}}\widehat{u}
\left( \frac{-t_{0,1}^{}}{2t_{0,2}^{}},\frac{-1}{4t_{0,2}^{}}\right),
\]
where $\widehat{u}$ as function of its arguments---not $\widehat{\varphi}$---
must satisfy the smoothness condition, the decay-at-spatial-infinity
condition, and the small norm assumption in order to guaranty the
applicability of the IST.
\medskip

{\bf Acknowledgments.} The author thanks A.~B.~Shabat and B.~G.~Konopelchenko
for the fruitful discussions. This work is supported in part by PRIN~97
``Sintesi'' and the Russian Foundation for Basic Research (Grant No.
99-01-00151).

\end{document}